\documentclass[a4paper, twocolumn]{article}

\usepackage{graphicx}
\usepackage{dcolumn}
\usepackage{bm}
\usepackage{siunitx}
\usepackage{hyperref}
\usepackage{comment}
\usepackage[utf8]{inputenc}
\usepackage[T1]{fontenc}
\usepackage{mathptmx}
\usepackage{etoolbox}
\usepackage{xcolor}
\usepackage[utf8]{inputenc}
\usepackage{authblk}

\usepackage[style=numeric, sorting=none, hyperref,
    url=false,
	doi=false,
	isbn=false,
	eprint=false,
	giveninits=true,
            sortcites=true,
            sorting=none,
            language=british]{biblatex}

\newbibmacro{string+doi}[1]{%
  \iffieldundef{doi}{%
    \iffieldundef{url}{#1}{\href{\thefield{url}}{#1}}
    }{%
    \href{http://dx.doi.org/\thefield{doi}}{#1}%
  }}
\DeclareFieldFormat{title}{\usebibmacro{string+doi}{\mkbibemph{#1}}}
\DeclareFieldFormat[article]{title}{\usebibmacro{string+doi}{\mkbibquote{#1}}}
\DeclareFieldFormat[thesis]{title}{\usebibmacro{string+doi}{\mkbibquote{#1}}}
\DeclareFieldFormat[inproceedings]{title}{\usebibmacro{string+doi}{\mkbibquote{#1}}}

\AtEveryBibitem{%
  \clearfield{note}%
	}

\definecolor{lapis}{HTML}{22577A}
\definecolor{hotpink}{HTML}{FF5733 }

\newcommand{\Rb}{Rb}
\newcommand{\esRb}{\textsuperscript{87}Rb}

\newcommand{\myfigref}[1]{Fig.~\ref{#1}}

\newcommand{\mysubfigref}[2]{Fig.~\hyperref[#1]{\ref*{#1}~(#2)}}

\makeatletter
\def\@email#1#2{%
 \endgroup
 \patchcmd{\titleblock@produce}
  {\frontmatter@RRAPformat}
  {\frontmatter@RRAPformat{\produce@RRAP{*#1\href{mailto:#2}{#2}}}\frontmatter@RRAPformat}
  {}{}
}%
\makeatother
\begin{document}


\title{Microstructured  optical fibres for quantum applications: perspective}
\author[1, *]{Cameron McGarry}
\author[1]{Kerrianne Harrington}
\author[1]{Alex O. C. Davis}
\author[1]{Peter J. Mosley}
\author[1]{Kristina R. Rusimova}

\affil[1]{Centre for Photonics and Photonic Materials, Department of Physics, University of Bath, Bath, BA2 7AY, UK}
\affil[*]{cdm34@bath.ac.uk}

\date{\today}

\maketitle

\textbf{
    Recent progress in the development and applications of microstructured optical fibres for quantum technologies is summarised. The optical nonlinearity of solid-core and gas-filled hollow-core fibres provides a valuable medium for the generation of quantum resource states, as well as for quantum frequency conversion between the operating wavelengths of existing quantum photonic material architectures. The low loss, low latency and low dispersion of hollow-core fibres make these fibres particularly attractive for both short- and long-distance links in quantum networks. Hollow-core fibres also promise to replace free-space optical components in a wide range of atomic experiments.
}

\section{\label{sec:intro} Introduction}

Quantum technologies promise to unleash unparalleled computational power, increased data security, and improved measurement precision; in the last decade alone we have witnessed some impressive advances in quantum computing,  cryptography, teleportation, metrology, and sensing. Light provides a uniquely attractive architecture for these emerging quantum technologies due to its high bandwidth, low decoherence, and the availability of methods to engineer its quantum properties such as entanglement and photon number statistics. The quantum technologies of the future are therefore likely to rely heavily on optical systems, especially for applications where quantum advantage has already been demonstrated, such as sensing, communications, and information processing. Long distance quantum communication and quantum cryptography could be achieved by distributing information between local quantum nodes, where information is generated, processed, and stored, via optical fibre networks. Schemes for on-demand photon generation, storage, switching, and multiplexing have also recently been developed and promise a route to overcoming some of the challenges posed by the need for high bandwidth, low loss, and fault tolerance. 

However, significant challenges remain in achieving seamless, low-loss, alignment-free integration between quantum network components and optical fibres. These problems are further exacerbated by the fact that there is no single wavelength that serves the needs of all quantum network functionalities – current photon sources, quantum memories, optical switches, quantum processors, and detectors cover a vast range of physical systems, which span the entire spectral range from the near-UV to the mid-IR. Previous attempts have focused on integrating on-chip architectures and atomic ensembles with the evanescent field of tapered nanofibres, or through grating couplers, edge couplers, and trenches. It has even been shown that it is possible to enhance these schemes by integrating them with in-fibre cavities. However, scalability of the free-space laser components in these systems remains an issue.

Microstructured optical fibres offer a promising route to overcoming some of these challenges. Unlike conventional optical fibres, where light is guided in a silica core through total internal reflection, the guidance properties of these fibres are dictated and can be tailored by the design of the microstructured glass cladding that surrounds the core. The fibre core itself can be either solid or hollow, with the structure of the cladding depending on the guidance mechanism \cite{Hecht21}. The guidance mechanism in solid-core photonic crystal fibre (PCF) is index guidance due to the effective index contrast between the solid core and the partially air-filled cladding. In hollow-core PCF, also known as photonic band gap fibre, guidance is instead attributed to a photonic band-gap effect\cite{Knight1998}. Anti-resonant hollow core fibres (HCF) have a simpler cladding structure that confines the light to the central core as the thin glass walls of the cladding create an anti-resonant effect that prevents light from escaping.

\begin{figure*}
    \includegraphics[width=1\textwidth]{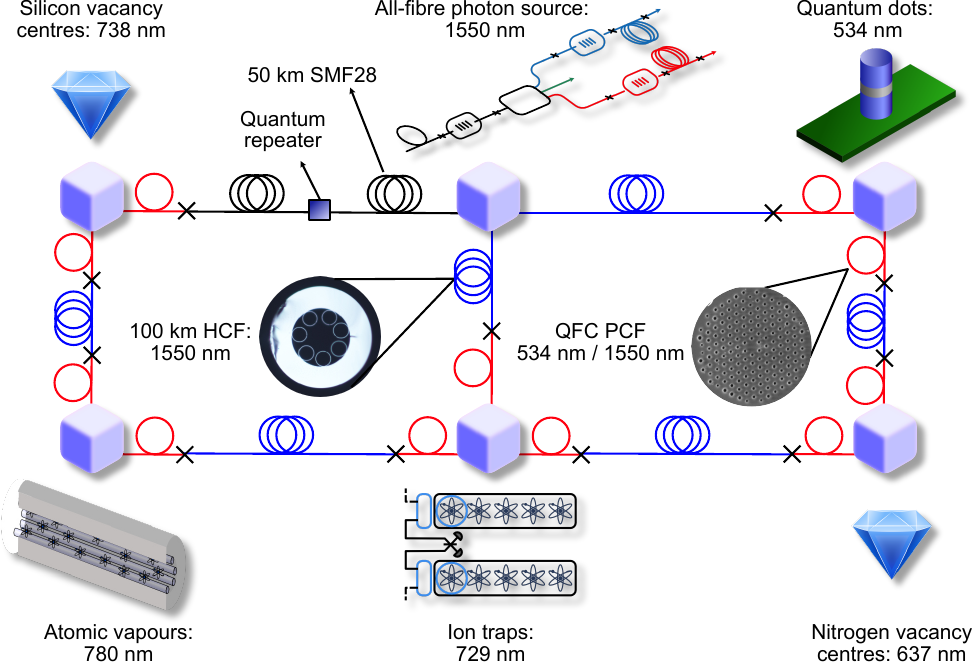}
\caption{\label{fig:master} A quantum network comprising nodes based on different quantum photonic architectures spanning a range of wavelengths and connected with all-fibre links.}
\end{figure*}

The variety of microstructured fibre brings a wide range of functionality for quantum technologies. By exploiting the nonlinearity of solid core PCF, it is possible to generate photon pairs directly inside optical fibres that can be seamlessly spliced to conventional solid-core networks. Similar fibres have also been demonstrated to achieve quantum frequency conversion, interfacing the wavelengths of interest for a broad range of quantum network components. Attenuation can be reduced by transmitting quantum information across mid- to long-range distances with HCF which offer fast, low-latency data transfer.  Coupling losses can be mitigated by integrating atomic vapours for storage and switching directly within the core of an HCF. Figure~\ref{fig:master} signposts some of the potential uses of microsctructured optical fibres in a possible future quantum network, where the operations performed at each node are based on different quantum architectures.

Here, we review recent progress in the use of microstructured optical fibres for quantum applications and outline some of the key challenges ahead of the field.

\section{\label{sec:sources} Sources and Multiplexing}

Correlated photon-pair generation enables the production of the resource states required for many photonic implementations of quantum technology. This includes both entangled pairs as well as the heralded single photons that can be derived from them. Quantum state generation is typically mediated by optical nonlinearity in the fibre core. Amorphous glasses generally do not possess the second-order ($\chi^{(2)}$) nonlinearity used in parametric down conversion sources, but fortunately the small mode areas and long interaction lengths in fibre permit significant third-order ($\chi^{(3)}$) interactions. PCF can offer a very high third-order nonlinearity typically in the range of \SIrange{10}{100}{\per\watt\per\kilo\meter}, exceeding highly nonlinear conventional fibres. This enables photon-pair generation in silica fibre by four-wave mixing (FWM), resulting in photon pairs that can be straightforwardly integrated into fibre networks. Meanwhile, HCF can be filled with an optically nonlinear gas with other desired properties such as Raman inactivity.

Through the design of the cladding, solid-core PCF enables much greater control over dispersion than conventional fibre. This is a result of the high index contrast between the air and the glass providing tight confinement of light, as well as the interplay between the mode shape and the ratio of air hole size, $d$, to air hole spacing or pitch, $\Lambda$. As a result, the fibre structure brings significant influence to bear on how the mode distribution, and therefore the transverse component of the wave vector, depends on wavelength, thus controlling the propagation constant. The wavelength range over which dispersion control via waveguide effects can compete with material dispersion encompasses conveniently many of the key wavelengths of interest for photonic quantum technologies, enabling photon-pair generation to be engineered in important areas of the spectrum. In addition, by keeping the ratio of hole diameter to pitch, $d/\Lambda$ below the critical value of 0.4 -- the so-called ``endlessly single-mode'' (ESM) condition -- ensures that the PCF supports only a fundamental mode across its entire transmission spectrum\cite{Birks1997Endlessly-single-mode-photonic}. Hence even highly nondegenerate photon pairs will be generated in the fundamental mode and can achieve high-efficiency coupling to conventional fibre, enabling PCF-based photon-pair sources to be fully fibre integrated\cite{McMillan2009Narrowband-high-fidelity-all-fibre} and interfaced with parametric downconversion (PDC) sources\cite{McMillan2013Two-photon-interference-between}.

These beneficial properties have driven the development of PCF-based photon-pair sources over the past 20 years. Early demonstrations confirmed the capacity of PCF to produce highly nondegenerate photon pairs, generated in ESM PCF pumped with a subnanosecond duration microchip laser\cite{Rarity2005Photonic-crystal-fiber} and in PCF with a high air-filling fraction pumped by a modelocked picosecond Ti:Sapphire laser\cite{Fulconis2005High-brightness-single}. Subsequently, tuning this Ti:Sapphire pump allowed the frequency dependence of the four-wave mixing (FWM) sidebands on the pump wavelength to be characterised to confirm the model of photon-pair generation subject to the PCF's dispersion\cite{Alibart2006Photon-pair-generation}. In 2007, the first demonstration of nonclassical Hong-Ou-Mandel interference (HOMI) between heralded single photons from independent PCF-based photon pair sources was achieved\cite{Fulconis2007Nonclassical-Interference-and-Entanglement} by building a duplicate of the source first reported by Fulconis \textit{et al}\cite{Fulconis2005High-brightness-single}. This is a critical capability both for building gates in linear optics quantum computation (LOQC) and for fusing photon pairs into triplets and larger cluster states for measurement-based quantum computation (MBQC). Additionally, in the same paper, the PCF was integrated into a Sagnac cavity in which bi-directional pumping combined with a 90 degree twist of the fibre axis enabled the direct generation of polarisation-entangled photon pairs. PCF has also been used to generate photon pairs when pumped by a GHz repetition rate vertical-external-cavity surface-emitting laser\cite{Morris2014Photon-pair-generation-in-photonic}, pointing towards the higher clock speeds desirable for photonic quantum computing.

Dispersion engineering in PCF enables more subtle control of the photon pair spectra, beyond simply producing nondegenerate wavelengths. In particular, many studies have focused on producing heralded single photons directly in pure quantum states. This requires the photon pair joint spectrum to be factorable into the product of one function dependent only on the signal frequency and another dependent only on the idler. This concept was first introduced in the context of PDC in nonlinear crystals\cite{Grice2001Eliminating-frequency-and-space-time} to remove the need for spectral filtering that is both lossy and detrimental to the photon number correlation of the signal and idler beams. However, similar conditions exist for FWM photon-pair generation in fibre; namely that if the group velocity of the pump is between those of the signal and idler, correlations between the signal and idler frequencies can be eliminated if the pump bandwidth is sufficiently broad. The heralding detection of one photon then yields no spectral information about its twin, allowing it to remain in a pure state capable of high-visibility HOMI\cite{Garay-Palmett2007Photon-pair-state-preparation, Cui2012Minimizing-the-frequency-correlation}. Such schemes have been implemented using birefringent phase matching in PCF\cite{Cohen2009Tailored-Photon-Pair-Generation, Halder2009Nonclassical-2-photon-interference, Clark2011Intrinsically-narrowband-pair} and with highly-nondegenerate signal and idler wavelengths \cite{Soller2010Bridging-visible-and-telecom, Cui2013Photonic-crystal-fibre-based}. The purity of the heralded photons in such schemes is limited firstly by the presence of side-lobes in the FWM phase matching function, though this may be addressed with a counterpropagating dual-pump configuration\cite{Fang2013}, and secondly by structural variations along the length of the fibre. Small changes in the PCF parameters introduced during fabrication can shift the phase matching by more than the FWM bandwidth, leading to beating between photon-pair generation amplitudes created at different points along the fibre, and causing the joint spectrum to break up\cite{Francis-Jones2016Characterisation-of-longitudinal-variation}. Such effects often limit the length of PCF that can be used to a few tens of centimetres, though it has been shown that rearranging different lengths of PCF can yield some control over the resulting unwanted spectral structure\cite{Cui2012Spectral-properties-of-photon}. In addition to producing high-purity heralded single photons, dispersion engineering in PCF also enables the generation of ultra-broadband photon pairs by creating flat group velocity dispersion over a wide wavelength range\cite{Garay-Palmett2008Ultrabroadband-photon-pair}.

An outstanding challenge is the spontaneous nature of photon-pair generation by both PDC and FWM, resulting in a superposition of photon number and probabilistic heralded single-photon generation. This shortcoming can be addressed by modification of the photon-number statistics through active routing of photons from several generation modes to a single master output conditioned on heralding detections. This technique is known as multiplexing and can be carried out using any degree of freedom in which light can be manipulated -- spatial and temporal modes, polarisation, and spectrum. Multiplexing has been implemented with sources based on a wide variety of photon-pair generation platforms. With FWM in PCF, all-fibre multiplexing of heralded single photons has been demonstrated in both the spatial\cite{Francis-Jones2016All-fiber-multiplexed-source} and temporal\cite{Hoggarth2017Resource-efficient-fibre-integrated-temporal} domains through addressing integration challenges including low-loss splicing between the PCF and conventional SMF, and isolation of the signal and idler guided modes from the pump. Once such a multiplexing network has been developed, it can also be used to reject uncorrelated noise from a single channel\cite{Francis-Jones2017Fibre-integrated-noise-gating}. 


HCF also provides unique opportunities for photon-pair generation. The very low nonlinearity of HCF is insufficient on its own for FWM. However, filling the core with one or more noble gasses at high pressure creates sufficient $\chi^\text{(3)}$ nonlinearity to generate photon pairs when pumped with an ultrafast laser. This brings the additional benefit that the noble gas is Raman inactive, eliminating one of the main causes of uncorrelated noise photons in FWM photon-pair sources\cite{Cordier2020Raman-free-fibered-photon-pair}. In addition, the HCF dispersion, and hence the photon-pair joint spectra, can be tailored by adjusting the gas pressure in the fibre, allowing spectral correlation to be actively controlled\cite{Cordier2019Active-engineering-of-four-wave}. The additional challenges ahead of this type of source -- creating a gas filled HCF and interfacing it with a fibre network -- are discussed in Section~\ref{sec:gascells}. The potential of liquid-filled HCF as a Raman-free FWM photon-pair source has also been investigated\cite{Barbier2015Spontaneous-four-wave-mixing, Cordier2017Raman-tailored-photonic-crystal}. HCF also provides the opportunity to move beyond the wavelengths of light that can be generated in solid-core PCF, which are of course limited by the transparency window of silica, further discussed in Section~\ref{sec:sources}. This is not the case in HCF, bringing the possibility of photon-pair generation at mid-IR wavelengths for quantum sensing, or even in the UV.

\section{\label{sec:QFC}Quantum frequency conversion}

In addition to photon-pair generation, FWM provides a convenient method to create photonic links between different wavelength bands. This is a key requirement for compatibility between processing nodes and memory technologies in future quantum networks.

Nonlinear quantum frequency conversion (QFC) has been investigated in a variety of platforms, through both second- and third-order nonlinearities. In fibre, typically QFC is achieved through Bragg-scattering FWM (BS-FWM), in which a source photon is converted to a target wavelength band when it scatters off a travelling periodic modulation of the refractive index created by two high-power pump fields. The frequency shift imparted to the source photon is the same as the detuning between the two pumps. As a result, BS-FWM is often suitable to impart smaller wavelength shifts than sum or difference frequency generation in second order nonlinear crystals. Again, the level of dispersion control available in PCF yields a high level of flexibility in the BS-FWM process, with tight modal confinement to the core boosting nonlinearity, and ESM behaviour accessible with suitable PCF cladding parameters.

It has been shown in theory that idealised QFC by BS-FWM in fibre allows unit conversion efficiency without noise from unwanted parametric processes\cite{McKinstrie2005Translation-of-quantum-states}. Inevitably, practical considerations such as higher-order dispersion, pulse walkoff, and Raman scattering can reduce conversion efficiency and introduce noise. Nevertheless it has been shown that careful optimisation of phase matching and pump power can restore near-optimal conversion efficiency\cite{Lefrancois2015Optimizing-optical-Bragg}, and judicious use of birefringence inhibits the phase matching of parasitic FWM processes\cite{Christensen2018Shape-preserving-and-unidirectional-frequency}. In addition, PCF can achieve an ultra-wide acceptance bandwidth for BS-FWM by engineering symmetry into the group velocity profile, providing a pathway to a universal quantum frequency interface with the telecoms C-band\cite{Parry2021Group-velocity-symmetry-in-photonic}.

In experiment, the first demonstration of QFC in PCF was published in 2005\cite{McGuinness2010Quantum-Frequency-Translation}, achieving almost 30\% conversion efficiency of heralded single photons between wavelengths of 683\,nm and 659\,nm. Subsequently, cascaded FWM processes in PCF -- seeded FWM followed by BS-FWM -- have allowed the number of ultrafast pump lasers to be reduced from two to one in a QFC link between 1092\,nm and 1550\,nm which achieved bi-directional conversion\cite{Wright2020Resource-efficient-frequency-conversion}. Recently, an ultra-broadband QFC scheme has been implemented in PCF with symmetric group velocity, converting heralded single photons from 1551\,nm to a tunable wavelength range between 1226\,nm and 1408\,nm in a single section of fibre\cite{Bonsma-Fisher2022Ultratunable-Quantum-Frequency}.

As a final note, due to its inherent flexibility, BS-FWM in PCF lends itself to applications beyond QFC. Examples include frequency-domain HOMI in which the BS-FWM process acts as a frequency beamsplitter to allow interference between photons of different colours\cite{Raymer2010Interference-of-two-photons-of-different}, with coherent photon conversion itself providing a route to carry out photonic quantum computation\cite{Langford2011Efficient-quantum-computing}.

\section{\label{sec:communications} Long distance communications}

Anti-resonant hollow core fibres hold much promise for multiple applications in quantum technologies, in addition to their use in long distance communications \cite{Nasti2022, Ding2020}. As quantum communication networks expand and increase in nodes, scaling up such systems will need some way of transmitting information over long distances \cite{DiAsamo2022}. For example, quantum information processing networks will need to scale up from on chip design, and move to several remote computational nodes interfaced with each other: an appropriate transmission medium will be needed \cite{Lu2021}. HCFs are one of the best candidates to offer a scaleable and robust solution for long-range quantum communication. 

Due to the hollow core, guided light interacts very little with the surrounding glass microstructure, promising low latency, low nonlinearity, and low loss at wavelengths not achievable in solid silica based fibres \cite{Jasion21, Mears2024}. Therefore, HCFs are particularly attractive for quantum network components which function with a variety of central wavelengths. For example, wavelengths between 580 nm and 900 nm are often of interest as many quantum systems use optical transitions that fall within this range \cite{Sakr2020}. It is therefore essential that candidates for a transmission medium for quantum applications can provide low loss for a very broad range of wavelengths to preserve information encoded in quantum states over long transmission lengths \cite{Nasti2022}. 

For long distance communications, the practical realisation of HCFs with lower attenuation than solid-core optical fibres particularly at telecommunications wavelengths has been a major fabrication challenge \cite{Jasion20, Jasion21}. Experiments with `NANF' (nested anti-resonant nodeless hollow core fiber) \cite{Jasion20}, and its descendent `DNANF' (double-nested), have been particularly successful, with reported lower loss in the C and L telecommunications bands than traditional low-loss solid core optical fibres \cite{Jaison22DNANF}. 

 Repeaterless or amplifierless ultra-broad-band transmission over HCF could be possible as lower and lower losses in fabricated fibres are achieved \cite{Poggiolini2022}. This is important in quantum applications because the use of amplifiers would cause decoherence of the quantum states destroying any quantum information. To reduce the cost overheads of quantum repeaters, ultra-low loss fibre is highly desirable \cite{Krutyanskiy2023}. To demonstrate record low loss over long transmission lengths, high yields of specialist HCFs are needed. The longest transmission line of low-loss HCF to date consisted of 7.72 km of fibre and was used to demonstrate data transmission in the C-band \cite{Nespola-21}. This transmission line was also used to demonstrate the first successful entangled photon distribution through long distance HCF, and highlighted the important properties of low latency and low dispersion for increased secure key rate in time-bin based QKD protocols \cite{Antesberger23}. The fabrication of high yields of such low-loss HCF remains a costly, but solvable, engineering challenge \cite{Jasion21}. As an alternative, splicing slightly different fibres, or re-circulating light into loops of fibre, is used to emulate longer lengths to overcome the practical challenges of fabricating the very high yields needed to truly test long range transmission in HCFs \cite{Nespola-21}. 

For other wavelengths of interest, comparable attenuation to silica fibres between 600 nm and 1100 nm has been achieved in HCFs. HCFs have demonstrated lower loss than silica at wavelengths of 850 nm, which have applications in short reach data centre links \cite{Sakr2020}, but also for caesium quantum memories that operate at 852 nm and 895 nm \cite{Reim2010}. Repeatedly producing the large yields needed here is less challenging than traditional telecommunications long distance links, as lengths are typically 100s of meters. The promised low thermal sensitivity and low dispersion features of HCFs have been demonstrated in real-time low-latency optically-switched intra-data center interconnects experimentally \cite{Zhou22, Zhou23}. Although HCFs have different physical dimensions to established solid-core fibre, such as core size and mode field diameter, progress has even been made for the efficient interconnection of fibres to established systems, such as vertical cavity surface emitting laser sources, used for data centres \cite{Jung2023}. Efficient interconnection is a necessary step for fully exploiting the advantages of HCFs.

In short, these efforts have shown that with the right fabrication techniques, it is realistic for HCFs to be lower loss than their older solid-core fibre counterparts at a wide variety of wavelengths, and that integration with existing systems is achievable.  These fibres are emerging as a suitable transmission medium for quantum communications and consequently have been used to for quantum key distribution (QKD) protocols and as entanglement distribution channels. An attractive feature for quantum communications is the ability to send co-propagating classical and quantum channels through the same fibre at different wavelengths. In HCFs, photons are less likely to be nonlinearly scattered between the quantum and classical channels \cite{Nasti2022}. Early experiments with HCFs have experimentally demonstrated discrete variable QKD, with record coexistence transmission of 1.6 tbps in the classical telecommunications channels \cite{Obada-2022} over a length of 2 km. For such applications, HCFs also offer better phase and thermal stability than other transmission media of interest \cite{Honz2023}. 

\section{\label{sec:CV}Squeezed and non-Gaussian state preparation}
PCF and gas-filled HCF also show considerable potential as sources of other nonclassical states for continuous-variable quantum information, sensing and communications applications. Squeezed states are a family of quantum states in which the Heisenberg uncertainty for some field quadrature is suppressed, or squeezed, below the threshold found in coherent (``semiclassical'') states\cite{schnabel2017squeezed}. 
They are an important resource across optical quantum technology, with applications in quantum sensing, networking and computing\cite{andersen201630}. There have been several demonstrations of squeezing that take advantage of the unique properties of microstructured fibres. Squeezed coherent states can be generated by self-phase modulation of bright laser pulses propagating down a nonlinear fibre. This so-called Kerr squeezing has been demonstrated in PCF around 800, 1310, and 1550 nm, with squeezing values up to 6.1 dB\cite{Fiorentino:02,lorenz2001squeezed,PhysRevLett.94.203601}. These states have been used to demonstrate measurements of optical transmissivity beyond the classical limit\cite{Atkinson2021}. Observation of a Kerr-type phase shift in PCF from signals at the single photon level has raised the prospect of a fibre-mediated controlled phase gate\cite{Matsuda2009}.

Bright two-mode squeezed states, known as ``twin beams'', can be generated by spontaneous four-wave mixing pumped in the high-gain regime, often with at least one of the fields being seeded with a coherent state. Typically this is realised with a single degenerate pump, with correlated signal and idler fields generated at equal frequency spacing from the pump. This has been accomplished in an argon-filled Kagome PCF, with the noise reduction factor\cite{ALLEVI2022127828} attaining a maximum of 1.7 dB and remaining measurable up to a mean photon number of 2500 per mode\cite{Chekhova:16}. 

Optical parametric oscillators (OPOs), where a nonlinear interaction is enhanced within a cavity, are versatile sources of both classical and non-classical light. Currently the best sources of single-mode squeezed vacuum are $\chi^{(2)}$-type OPOs\cite{schnabel2017squeezed}. Microstructured fibres of various types have been investigated for improving coupling efficiency for these types of sources\cite{Brieussel:18}. There have been a number of demonstrations of a $\chi^{(3)}$-type OPO with ring-cavity geometry using PCF for classical applications\cite{van2011fiber}, but the approaches taken to achieve optical feedback, such as splicing to close the loop, introduces significant losses which limit their usefulness as quantum sources. Next-generation fibre-integrated squeezed vacuum sources consisting of a longitudinal Bragg-reflector cavity inside a section of PCF, forming a $\chi^{(3)}$-type OPO, are currently in development\cite{davis2023squeezed}. 

\begin{figure}
\includegraphics[width=\linewidth]
{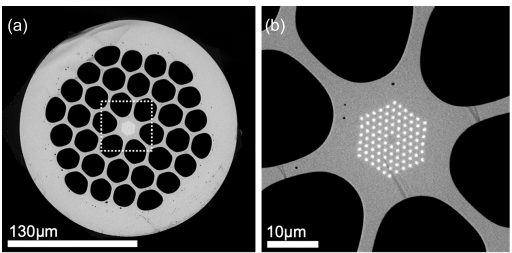}
\caption{\label{fig:tripletpcf}
The flexibility in microstructure design can allow extreme dispersion engineering. This hybrid PCF is phase matched for spontaneous photon triplet generation (or third harmonic generation) between 520 nm and 1560 nm. Reproduced under the CC BY license from Cavanna~et~al.~\cite{Cavanna2020}}
\end{figure}

Beyond squeezing, microstructured fibres are perhaps the most promising platform for spontaneous photon triplet generation (STRIG) via third-order parametric down conversion. STRIG is a longstanding goal as it allows the deterministic generation of non-Gaussian states without the need for postselection, offering a route to scalable quantum computation. With only a single pump field, the third-order interaction is extremely weak, requiring very high (sometimes called ``giant'') optical nonlinearity. Another key challenge is phase-matching the pump and signal wavelengths, which differ by a factor of 3, necessitating extreme dispersion engineering. A variety of approaches have been studied including xenon-filled HCF, tapered fibre, and hybrid PCF\cite{Cavanna2020} (see \myfigref{fig:tripletpcf}). So far STRIG is yet to be observed, but proposals such as longitudinal cavities and using different materials suggest a path towards this important milestone.

\section{\label{sec:gascells} Fibre gas cells}

\begin{figure*}
    \includegraphics[width=\textwidth]{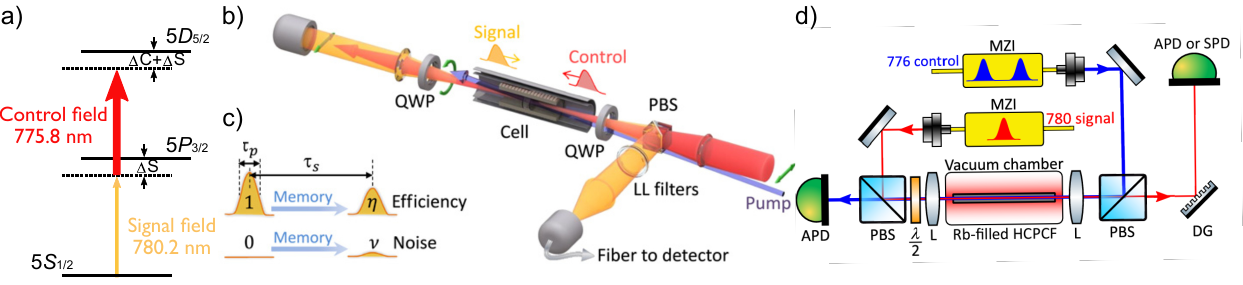}
\caption{\label{fig:memory}
The fast ladder memory scheme~\cite{Finkelstein2018} in \esRb{}. (a) The relevant \esRb{} structure. (b) Counter-propagating signal (yellow) and control (red) beams are incident on a macroscopic \Rb{} vapour cell. This allows the storage of the signal pulse in the collective excitation of the atomic vapour. (c) The pulse is stored for some time $\tau_s$ and retrieved with an efficiency $\eta$ by the application of a second control pulse. (d) Schematic of implementation of the memory scheme in a fibre-integrated vapour cell (reproduced with permission from Rowland~et~al.~\cite{Rowland2024}). Parts (a-c) reproduced and adapted under the CC BY-NC license from Finkelstein~et~al.~\cite{Finkelstein2018}.
}
\end{figure*}

We have already discussed the concept of a gas-filled HCF as a single-photon source, for QCF and for generating non-classical states. 
In addition to these use-cases, a PCF or HCF filled with a vapour could be used to replace conventional macroscopic free space vapour cells, which are used in a wide range of atomic experiments.
This integration of the vapour cell directly into the fibre network can allow for reduction of coupling losses, enhancing the fidelity of the entire system; as well as providing a route to miniaturisation and scalability.
Additionally, it significantly reduces the mode area, and potentially increases the interaction length, allowing much stronger interaction between the light and the atoms.

In particular, we will discuss the applications of a fibre gas cell for optical memories and quantum sensing.
An optical memory scheme, such as the Fast Ladder Memory~\cite{Finkelstein2018} depicted in \myfigref{fig:memory}, stores a signal field in a collective spin excitation of a warm atomic vapour by mapping to a higher state via a strong control field, which is close to resonance with a two-photon transition to a higher state.
The pulse can be retrieved at a later time by re-application of the control field.
The atom of choice is usually an alakali metal, in particular \esRb{} is favourable due to Doppler cancellation of the signal and control fields (due to their proximity in wavelength) which extends the memory lifetime.

Integration of atomic vapour cells into fibre will therefore permit the implementation of these same memory schemes directly within the fibre network, allowing an additional method of multiplexing, alongside those described in section~\ref{sec:sources}.
Combined with existing schemes for spatial multiplexing (also described in section~\ref{sec:sources}), temporal multiplexing will result in powerful tools for single-photon operation entirely in fibre; there are also applications in the generation of large resource states, with applications in MBQC~\cite{Asavanant2019}.

Additionally, alkali vapours are powerful tools for quantum sensing, in particular by the creation of Rydberg atoms~\cite{Adams2020}.
These highly excited atoms have extremely large electric dipole moments, so they interact strongly with electric fields.
This can be harnessed for applications in quantum computing~\cite{Saffman2010}, such as by long-range interaction between sites of an atomic lattice, or utilised in the quantum sensing of external fields.
Embedding Rydberg atoms within a microstructured fibre could therefore yield a powerful and portable sensor of external fields such as microwaves~\cite{Simons2018, Menchetti2023}.

%

Gas-filled HCFs have been previously used to implement in-fibre gas lasers~\cite{Hassan2016} and  optical gas sensors~\cite{Warrington2023, Suslov2023}.
However, loading with alkalis for quantum applications presents a much greater engineering challenge, due to their high reactivity with atmospheric water.
Various experiments with atoms in fibres have demonstrated that loading~\cite{Epple2014, Bajcsy2011}, spectroscopy~\cite{Xin2018} and memories~\cite{Sprague2014} can be performed in microstructured fibres.
However, these are all with the limitation that they are performed in vacuum chambers, with no means of connectorising the filled HCF and attaching to a large fibre system.
Additionally, there is often a reliance on extensive laser cooling systems to facilitate the loading of atoms into the fibre.

In the rest of this section, we will summarise the outlook for fibre gas cells, with a focus on two key challenges facing their development.
In the case of memories, the interconnections must also have incredibly low insertion loss to the opposing fibre.
This is essential for mitigating the loss of information-carrying photons, and maximising system fidelity.
This first challenge is discussed in detail in~\cite{McGarry2024}, and will be outlined further in the next section.
The second challenge is that of filling the fibre with the vapour; although the process of filling a macroscopic cell is well understood, the same cannot be said for these much smaller, fibre-integrated cells.
There is an open question of whether the HCF can be loaded before encapsulation, or if this must be done after the fact -- in which case it will be necessary to create holes in the fibre walls through which filling can take place.

We will also address an ancillary challenge: the diameter of the HCF core is much smaller than that of a vapour cell.
Hence, we can expect a much higher rate of collisions between the atoms and the walls, which will lead to increased rate of atomic decoherence due to spin relaxation.
We will outline ways in which this effect can be mitigated with anti spin-relaxation coatings, or entirely avoided by using phase modulation schemes rather than memory schemes.

\subsection{\label{sec:connecting} Connectorising}

\begin{figure*}
    \includegraphics[width=1\textwidth]{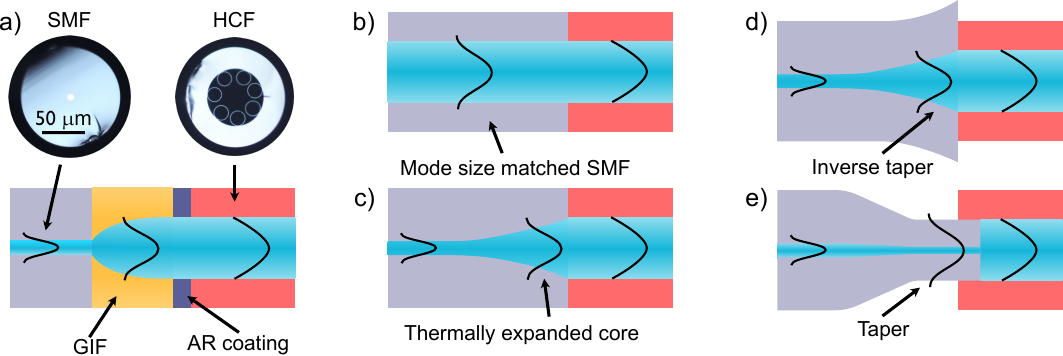}
\caption{\label{fig:connectorising} Low-loss connections between HCF and solid-core SMF. (a) Schematic of an SMF-GIF-HCF connection. Inset: optical microscope images of SMF and HCF cross-sections on the same scale. Schematics of low-loss SMF-HCF connections through (b) mode sized matched SMF, (c) SMF with a thermally expanded core, (d) inverse-tapered SMF, and (e) adiabatically tapered SMF. (b)-(e) adapted under the CC BY license from Slav\'ik~et~al.~\cite{Slavik2023}.}
\end{figure*}

As described previously, for an HCF cell (or indeed any optical device) to be of use in an optical quantum system, it must have sufficiently low insertion loss.
This is to prevent the loss of single photons carrying the quantum information.
Any such losses will have a significant impact on the efficiency of the system, requiring additional redundancy and quantum error correction~\cite{Pankovich2023}.

The challenge here is that the HCF cell will typically need to connect to a conventional SMF for transport of the photonic qubit, and these two fibres have significantly different spatial modes.
The fundamental mode of an HCF can be assumed to have the spatial profile of a truncated Bessel function, with circular symmetry~\cite{Murphy2023}, whereas an SMF mode has the usual form of an $\text{LP}_{0,1}$ mode~\cite{SandL}.
The spatial range of these modes is determined by the size of the core, which for SMF is on the order of a micron, and for HCF is on the order of tens of microns, as can be seen in \mysubfigref{fig:connectorising}{a}.
There is therefore a significant mode mismatch which must be overcome to achieve a highly efficient coupling.

Fortunately, low-loss interconnects between SMF and HCF are a developing technology, and there are already a number of very promising studies into this field.
Chief among these are the works of the University of Southampton Optical Research Centre, who have pioneered low-loss interconnections between SMF and HCF via a graded index-lensing technique~\cite{Suslov2023, Slavik2023}.
In this scheme, a short segment of graded index fibre (GIF) is fusion spliced onto the SMF.
The length of the GIF is chosen so that the light pattern which exits is expanded (and typically, collimated) so that this mode more closely matches the fundamental mode of the HCF.
This scheme is pictured in \mysubfigref{fig:connectorising}{a}, and has been employed to achieve extremely low insertion losses of \SI{0.15}{\decibel}~\cite{Suslov2021}.

It is also possible to perform so-called gap-engineering, where the distance between the GIF and HCF is non-zero, and so the light propagates to match the HCF mode.
We will discuss below the significance of this effect for filling of HCFs.
The details of these low-loss interconnects represent a rich field, and a recent review on the subject explores these in greater detail~\cite{Slavik2023}.

HCF-GIF-SMF interconnects can be used to build a low-insertion-loss fibre
cell. Insertion loss of \SI{0.6+-2}{\decibel} for a complete cell has been achieved at a wavelength of
\SI{780}{\nano\meter}~\cite{McGarry2024}, compatible with the aforementioned Rb
memory schemes. However, it should be noted that the coupling efficiency of
these interconnects is fundamentally limited by the differing shapes of the SMF
and HCF modes. A GIF lens does not change the overall shape of the mode, it only
alters its width, and so the fundamental limit is
\SI{0.07}{\decibel}~\cite{Suslov2021}. Other interconnection techniques will
have to be developed to go below this limit. This could include some combination
of the above, along with: mode-matching directly with the SMF, thermally
expanding the SMF core, an inverse taper~\cite{Kakarantzas2007}, or a
standard taper~\cite{Birks1992, Pennetta2017}, as depicted in
\mysubfigref{fig:connectorising}{b-e} respectively.

\subsection{\label{sec:switching} Fibre gas filling}

Fibre gas filling remains a challenge for HCF, where filling time primarily depends on length and core size.
Pressurization chambers for controllable filling of HCFs with different gases, whilst allowing for optical excitation, are well established \cite{love-2018}.

%
%

%
Fibres have been successfully gas filled for a number of applications such as fibre lasers \cite{Pryamikov2022}, based on absorption of gas molecules in the hollow core causing population inversion \cite{XuMengrong2018, Aghbolagh19}, stimulated Raman scattering or resonant dispersive waves \cite{Sabbah2024} sensing \cite{Suslov2023}, and post-processing \cite{Yerolatsitis17}. However, efficient gas filling of long lengths of HCF is still a challenge \cite{Masum2020, Warrington2023}, especially for filling with vapours like rubidium~\cite{Rowland2024} where filled fibres disconnected from a vacuum chamber do not yet exist.

An alternative approach would be to load the fibres with rubidium atoms through holes in the side walls.
Previous work has shown that holes can be ablated in the side of HCF, allowing for the introduction of gases for sensing~\cite{Warrington2023}.
A similar approach could be used to fill HCF with rubidium, for example by first sealing the micro-drilled holes inside a vapour cell, which can then be filled by conventional methods. 
In this case the fibre ends would be sealed by low-loss interconnects so as to prevent the vapour escaping.
This would also simplify connectorisation, as there would be no need to do this in a controlled environment.
The rubidium would diffuse into the fibre over the timescale of many days or weeks. However many fibres could be filled at once by this method, creating a bank of vapour cells for use in multiplexing.

\subsection{Use in photonic quantum technologies}

Once realised, a fibre-integrated vapour cell can be used analogously to existing macroscopic vapour cells, for example to implment optical quantum memories as outlined above.
Low-loss interconnects will enable increased efficiency due to the reduced coupling loss in the macroscopic experiment.
Further, as we stated above, the reduced mode area inside the fibre will allow for stronger light-matter interactions, reducing demand on laser systems used in the memory schemes.
This comes at a cost: the small core size means the atoms are much more tightly confined, so as they move with their thermal drift, they will collide with the walls more often than in a macroscopic cell.
This will increase the rate of spin-relaxation, and hence reduce memory lifetimes.

This can be mitigated by the application of anti spin-relaxation coatings to the interior of the HCF.
These are commonly used in conventional vapour cells, and have been previously demonstrated not to significantly alter the guidance properties of the fibre~\cite{Rusimova2019, Bradley2013}.
Alternatively, these spin relaxations can be avoided entirely by relying not on an optical memory, but on all-optical phase modulation~\cite{Davis2023}.
This can also be used to enable multiplexing, but here the process occurs on the timescale of the signal pulse, removing the need for prolonged coherence of the atomic states.

\section{\label{sec:conclusions} Conclusions}
Microsctructured optical fibres are a useful medium for both short-range and long-range quantum network components. Even though technical challenges still remain around interfacing optical fibres with the existing quantum platforms, progress in this area of research has been rapid in the last few years. PCF and HCF have already shown their utility as sources and mode converters of quantum light, as well as interconnectors between quantum nodes. We anticipate that in the near-term these systems will play a key role in future quantum networks and other quantum technologies.

\section*{Acknowledgements}

CMcG, PJM, and KRR wish to acknowledge the support of Innovate UK Quantum Data Centre of the Future grant 10004793. AOCD acknowledges funding from UKRI EPSRC grant ``PHOCIS: A Photonic Crystal Integrated Squeezer'' EP/W028336/1. PJM acknowledges support from the UK Hub in Quantum Computing and Simulation, part of the UK National Quantum Technologies Programme with funding from UKRI EPSRC (EP/T001062/1). KH acknowledges support from UKRI EPSRC (EP/T020903/1). KRR acknowledges support from the Royal Society (RGS/R1/231369) and from UKRI EPSRC (EP/X031934/1).

\section*{License}

The above article has been submitted to APL Quantum. After it is published, it will be found at \url{https://pubs.aip.org/aip/apq}.

Copyright 2024, McGarry et al. This article is distributed under a Creative Commons Attribution-NonCommercial 4.0 International (CC BY-NC) License. \url{https://creativecommons.org/licenses/by-nc/4.0/}

Copyright 2024, McGarry et al. This article is distributed under a Creative Commons Attribution-NonCommercial-NoDerivs 4.0 International (CC BY-NC-ND) License. \url{https://creativecommons.org/licenses/by-nc-nd/4.0/}

\printbibliography

\end{document}
%